\documentclass[pss]{wiley2sp} 
\usepackage{amsmath}
\usepackage{citesort}

\begin{document}

\title{Beyond the quantum spin liquid concept in frustrated 
two dimensional organic superconductors}


\author{%
  R.T. Clay\textsuperscript{\Ast,\textsf{\bfseries 1,2}},
  S. Dayal\textsuperscript{\textsf{\bfseries 1}},
  H. Li\textsuperscript{\textsf{\bfseries 3}}, S. Mazumdar\textsuperscript{\textsf{\bfseries 3}}}

\authorrunning{R.T. Clay et al.}

\mail{e-mail
  \textsf{r.t.clay@msstate.edu}, Phone:
  +1-662-325-0628, Fax: +1-662-325-8898}

\institute{%
  \textsuperscript{1}\,Department of Physics and Astronomy and HPC$^2$ Center for 
Computational Sciences, Mississippi State University, Mississippi State, MS 39762, USA\\
  \textsuperscript{2}\,Institute for Solid State Physics, The University of Tokyo,
Kashiwa 277-8581, Japan\\
  \textsuperscript{3}\,Department of Physics, University of Arizona, Tucson,
AZ 85721, USA}

\received{XXXX, revised XXXX, accepted XXXX}
\published{XXXX}


\abstract{ \abstcol{The occurrence of antiferromagnetism in
    $\kappa$-(ET)$_2$X can be understood within an effective
    $\frac{1}{2}$-filled band with dimers of ET molecules containing
    one hole each.  We argue that while this effective model can
    describe the presence of antiferromagnetism, a complete
    description for these materials requires the correct carrier
    density of one-half per molecule. For dimerized and strongly
    frustrated $\frac{1}{4}$-filled lattices we show that a
    singlet-paired state}{coexisting with charge ordering occurs that
    we have termed the Paired Electron Crystal (PEC). Here we
    investigate the $\frac{1}{4}$-filled model on a dimerized lattice,
    showing regions where AFM, PEC, and the Wigner-crystal occur.  We
    point out the need to go beyond quantum spin liquid concepts for
    highly frustrated materials such as
    $\kappa$-(ET)$_2$Cu$_2$(CN)$_3$ and
    $\beta^\prime$-EtMe$_3$Sb[Pd(dmit)$_2$]$_2$ which we believe are
    PECs at low temperatures.}}

\maketitle   

\section{Introduction}

A challenging problem in condensed matter physics is to describe
systems where both electron-electron (e-e) interactions and lattice
frustration are significant.
The possibility that on a
frustrated lattice a quantum spin system may remain disordered even at
low temperature forming a quantum spin liquid (QSL), has spurred much
theoretical and experimental work \cite{Balents10a}.  
Two
candidate QSL materials, $\kappa$-(ET)$_2$Cu$_2$(CN)$_3$ (hereafter
$\kappa$-CN) and $\beta^\prime$-EtMe$_3$Sb[Pd(dmit)$_2$]$_2$
(hereafter dmit-Sb), have been identified from within the group of
charge transfer solid (CTS) organic
superconductors \cite{Kanoda11a}. In addition to these particular two
compounds displaying QSL-like behavior, the $\kappa$-(ET)$_2$ and
$\beta^\prime$-X[Pd(dmit)$_2$]$_2$ series are of wide interest because
they display other exotic phases including antiferromagnetism (AFM), a
so-called valence-bond solid (VBS) phase, and superconductivity (SC)
\cite{Dressel11a,Kato04a}.

Structurally both $\kappa$-CN and dmit-Sb consist of two-dimensional
layers of dimers arranged in an anisotropic triangular lattice. With
an average of 0.5 carriers per molecule this has led to the common
description of both within an effective $\frac{1}{2}$-filled band
where each dimer is replaced by a single site. In the presence of
strong on-site e-e interactions one obtains a quantum spin model where
each spin represents a pair of molecules \cite{Kanoda11a}.

The $\frac{1}{2}$-filled Hubbard model on an anisotropic triangular
lattice is frequently taken as a minimal electronic model for these
materials \cite{Powell11a}.  This model contains a single frustrating
bond $t^\prime$ per square plaquette. In the case of weak anisotropy
(small $t^\prime/t$), the ground state has Ne\'el antiferromagnetic
order. As $t^\prime/t$ is increased, frustration of the AFM leads to a
paramagnetic metallic (PM) state.  Based on mean-field calculations
many investigators have suggested that a region of $d_{x^2-y^2}$ SC
mediated by spin fluctuations emerges near the boundary between the
AFM and PM phases. This proposal ignores other CTS where the
superconducting state is proximate not to the AFM but to charge
ordered or VBS phases.  More importantly, calculations beyond the
mean-field approximation do not find SC within the
$\frac{1}{2}$-filled model for any $U$ and $t^\prime/t$
\cite{Clay08a,Tocchio09a}.  Our earlier exact calculations were for a
finite 4$\times$4 cluster \cite{Clay08a}.  More recently, we have
adopted the Path Integral Renormalization Group method
\cite{Kashima01b} to perform calculations for much larger lattices
confirming the absence of SC \cite{Dayal11b}.

The absence of SC within the $\frac{1}{2}$-filled effective model for
the CTS suggests that one must return to the underlying
$\frac{1}{4}$-filled lattice in order to fully understand the
electronic properties.  We have recently shown that in the
$\frac{1}{4}$-filled band, strong e-e correlations and lattice
frustration lead to a transition from AFM to a paired charge-ordered
state, the Paired Electron Crystal (PEC) \cite{Li10a,Dayal11a}. The
PEC can be visualized as the $\frac{1}{4}$-filled equivalent of a VBS
state in a quantum spin model. In the $\frac{1}{4}$-filled picture,
each single spin of the spin model must be replaced by a pair (dimer)
of sites. Once singlet bonds are formed between dimers, the charge
occupation within each dimer necessarily becomes nonuniform---unlike
the $\frac{1}{2}$-filled band, at $\frac{1}{4}$-filling charge and
spin degrees of freedom are linked.  Interestingly, recent experiments
(discussed below in Section 3) have found evidence for charge
disproportionation in the optical spectra and the frequency dependence
of the dielectric response at low temperature.

In references \cite{Li10a,Dayal11a} we considered a
$\frac{1}{4}$-filled an-isotropic triangular lattice with Coulomb and
electron-phonon (e-p) interactions where along any one direction ($x$, $y$
or $x+y$) all bonds were taken to be uniform in the absence of
instabilities.  In order to simulate the dimerized $\kappa$ or dmit
lattice, we have subsequently repeated these calculations for a
lattice in which the sites have formed rigid dimer units along one
direction. We report these results here.  The organization of the
paper is as follows. In Section \ref{results} we describe calculations
showing that the PEC state occurs also within the lattice of coupled
dimers. In Section \ref{discussion} we discuss recent experiments for
the proposed CTS QSL materials, that suggest that at the lowest
temperatures a $\frac{1}{4}$-filled model is essential to correctly
describe their electronic properties.

\section{Dimer lattice calculations}
\label{results}

\begin{figure}
  \centerline{\resizebox{3.2in}{!}{\includegraphics{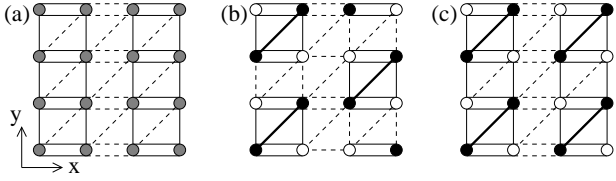}}}
 \caption{(a) Dimer lattice considered here. Double lines correspond to
   rigid bonds with hopping matrix elements $t+\delta t$.  Dashed and
   double dashed lines have hopping matrix elements of $t-\delta t$ in
   the absence of lattice distortion.  Electron-phonon coupling is
   included on inter-dimer bonds in $x$ and $y$ directions. (b) PEC
   state. Here filled (empty) sites have charge density of 0.5+$\delta$
   (0.5-$\delta$). (c) The WC-SG state. In (b) and (c), heavy lines
   indicate singlet bonds.}
 \label{lattices}
\end{figure}

\begin{figure}
\centerline{\resizebox{3.0in}{!}{\includegraphics{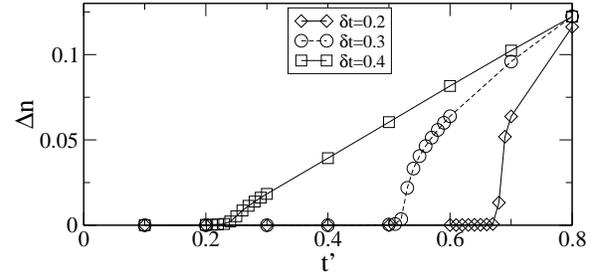}}}
\caption{ Charge disproportionation $\Delta n$ in the dimer lattice
  versus $t^\prime$.  Parameters are $U=6$, $V_x=V_y=1$, $V^\prime=0$,
  $\alpha=1.0$, $\beta=0.1$, $K_\alpha=K_\beta=2$. Diamonds, circles,
  and squares are for $x$-axis dimerization of $\delta t=$0.2, 0.3,
  and 0.4, respectively.  Lines are guides to the eye.}
\label{deltan}
\end{figure}
We consider the following Hamiltonian,
\begin{eqnarray}
H&=&-\sum_{\nu,\langle ij\rangle_\nu}t_\nu(1+\alpha_\nu\Delta_{ij})B_{ij} \label{ham} 
+\frac{1}{2}\sum_{\nu,\langle ij\rangle_\nu} K^\nu_\alpha \Delta_{ij}^2 \\
&+&\beta \sum_i v_i n_i + \frac{1}{2}K_\beta \sum_i v_i^2  \nonumber \\
&+& U\sum_i n_{i\uparrow}n_{i\downarrow} + 
\frac{1}{2}\sum_{\langle ij\rangle}V_{ij} n_i n_j. \nonumber 
\end{eqnarray}
In Eq.~\ref{ham},
$B_{ij}=\sum_\sigma(c^\dagger_{i,\sigma}c_{j,\sigma}+H.c.)$ is the
kinetic energy operator between sites $i$ and $j$. The index $\nu$
refers to the lattice directions $x$, $y$, and $x+y$ (see
Fig.~\ref{lattices}).  $\alpha_\nu$ is the inter-site 
e-p coupling for direction $\nu$, with corresponding spring constant
$K_\nu$.  $\beta$ is the onsite (Holstein) e-p coupling with
corresponding spring constant $K_\beta$. $U$ and $V$ are respectively
the onsite and nearest-neighbor Coulomb interactions.  We have
previously considered the case where the intersite e-p coupling
applies to all $t_x$ and $t_y$ bonds \cite{Li10a,Dayal11a}. Here we
will consider a lattice consisting of rigidly bound dimers connected
by deformable bonds, with frustrating bonds $t^\prime\equiv t_{x+y}$
across each plaquette of the lattice as shown in
Fig.~\ref{lattices}(a).  In this lattice $t_x$ varies as $t_x +
(-1)^{i_x}\delta t$, where $i_x$ is a site index in the $x$
direction. The bond modulation induced by the e-p coupling
$\Delta_{i_x,i_x+1}$ is set to zero for the strong intra-dimer bonds,
and is allowed to vary self-consistently for the inter-dimer bonds
along the $x$-direction.  E-p coupling along the $y$ axis
is not restricted. As in reference \cite{Dayal11a}, we take
$\alpha_x=\alpha_y$, $K_x=K_y=K_\beta=2$, $t_x=t_y\equiv t$, and
measure energies in units of $t$.  We solve Eq.~\ref{ham} exactly on a
$4\times4$ lattice.

As in the lattice without rigid dimerization \cite{Li10a,Dayal11a},
when the strength of the frustrating bond $t^\prime$ is increased we
find a transition from AFM (Fig.~\ref{lattices}(a) with uniform charge
density on each site) to a charge-disproportionated PEC state
(Fig.~\ref{lattices}(b)). Fig.~\ref{deltan} shows the charge
disproportionation $\Delta n$, defined here as the charge density
difference between sites with large and small charge densities, as a
function of $t^\prime$.  For $t^\prime>t^\prime_c$, $\Delta n$ becomes
non-zero indicating charge disproportionation within each
dimer. Simultaneous loss of antiferromagnetic order and the opening of
a spin gap due to nearest-neighbor singlet formation as indicated in
Fig.~\ref{lattices}(b) is seen in the bond orders and spin-spin
correlations \cite{Li10a,Dayal11a} (not shown here).  Assuming equal
e-p coupling strength, we find that stronger dimerization (larger
$\delta t$) gives the PEC state at a {\it weaker} level of lattice
frustration.  However, as is seen in Fig.~\ref{deltan}, for strong
dimerization, $\Delta n$ remains relatively small until $t^\prime$ is
significantly larger than $t^\prime_c$. This may have experimental
implications.

\begin{figure}
\centerline{\resizebox{3.0in}{!}{\includegraphics{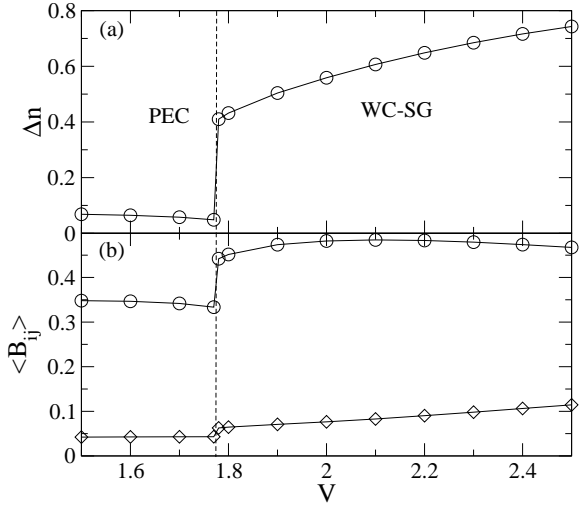}}}
\caption{Variation of (a) charge disproportionation, $\Delta n$,
and (b)  bond order between two successive bonds in the $x+y$ direction
with $V$. Nearest-neighbor Coulomb interactions of the form $V_x=V_y=V$
and $V^\prime=0$ are assumed. Other parameters are $U=6$, $t^\prime=0.6$,
$\alpha=1.0$, $\beta=0.1$, $K_\alpha=K_\beta=2$. Lines are
guides to the eye.}
\label{pec-wcsg}
\end{figure}

\begin{figure}
\centerline{\resizebox{3.0in}{!}{\includegraphics{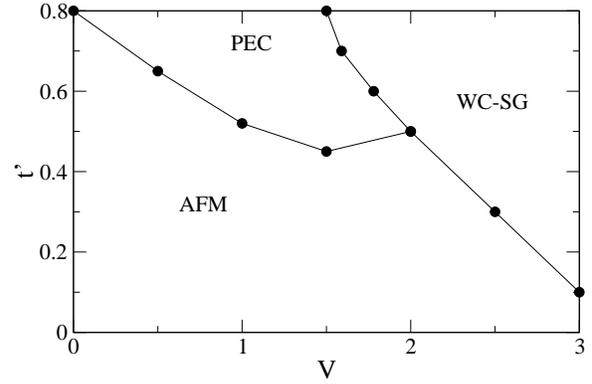}}}
\caption{Ground state phase diagram of the dimer model for $\delta
  t=0.3$, $U=6$, $V_x=V_y=V$, and $V^\prime=0$.  Lines are guides to
  the eye.}
\label{phasediag}
\end{figure}

When nearest-neighbor Coulomb interactions $V_{ij}$ in Eq.~\ref{ham}
are stronger, the checkerboard-pattern Wigner Crystal (WC) state shown
in Fig.~\ref{lattices}(c) can become more stable than the PEC of
Fig.~\ref{lattices}(b). Here we choose $V_{ij}$ to be of the form
$V_x=V_y\equiv V$ but $V^\prime=0$. Other choices of $V_{ij}$ can
change the stability of the WC significantly---for example taking
$V^\prime>0$ suppresses the checkerboard WC shown in
Fig.~\ref{lattices}(c) and stabilizes the PEC.  \cite{Dayal11a}.  As
shown in Fig.~\ref{pec-wcsg}(a), $\Delta n$ in the WC state is
considerably larger than in the PEC state. The charge and bond
distortion of the WC state is shown in Fig.~\ref{lattices}(c). Because
of the rigid dimers in this lattice, the bond strength alternates
strong-weak-strong along the $x+y$ ($t^\prime$) direction. The bond
alternation in the one dimensional chains along the $x+y$ direction
gives this phase a spin gap, denoted as ``WC-SG'' in reference
\cite{Dayal11a}.  In Fig.~\ref{phasediag} we show the $t^\prime-V$
ground state phase diagram.

Summarizing our numerical results for the case the lattice of rigid
dimers: (i) As in the lattice with unconstrained $t_x$ bonds we find
an AFM-PEC transition at a critical frustration $t^\prime_c$. Stronger
dimerization leads to PEC formation over a larger region of phase
space, however with a weaker $\Delta n$ close to $t^\prime_c$.  (ii)
When nearest-neighbor Coulomb interactions are significant a WC phase
also occurs.  In the case of a rigid dimer lattice this WC phase has a
much greater tendency towards spin gap.

\section{CTS spin liquid candidates}
\label{discussion}

In this section we discuss $\kappa$-CN and dmit-Sb in detail. SC
occurs in both: in $\kappa$-CN under pressure \cite{Komatsu96a}, and
in other dmit materials \cite{Shimizu07a}.  For both materials a
number of recent experiments indicate transitions to broken-symmetry
states at low temperature.

\subsection{$\kappa$-(ET)$_2$Cu$_2$(CN)$_3$}

In $\kappa$-CN the estimated Heisenberg exchange integral between
nearest-neighbor dimers is $J\sim$ 220--250 K
\cite{Shimizu03a}. Despite the significant magnetic interaction
between dimers $^1$H NMR experiments find no long range magnetic order
to mK temperatures \cite{Shimizu03a}.  However, a number of
experiments suggest that a second order phase transition occurs at
$\sim$ 6 K
\cite{Shimizu06a,Yamashita08a,Yamashita09a,AbdelJawad10a,Manna10a,Pratt11a}.
Below 6K the $^{13}$C NMR relaxation rate 1/T$_1$ decreases,
suggesting a spin gapped ground state \cite{Shimizu06a}.  A decrease
in the spin fluctuation rate is also seen below 6K in zero-field
$\mu$SR \cite{Pratt11a}.  Field-dependent $\mu$SR finds a critical
field $H_c$ above which the linewidth increases rapidly and estimate
the spin gap as 3.5 mK \cite{Pratt11a}.  Precision measurement of
lattice expansion finds strong lattice effects suggesting a
second-order phase transition at 6K \cite{Manna10a}.  The dielectric
response is frequency dependent below 60 K suggesting an
antiferroelectric ordering of dipoles below 6 K, requiring unequal
charge densities within each dimer \cite{AbdelJawad10a}.  One of the
most puzzling aspects of is that while certain measurements ($^{13}$C
NMR, $\mu$SR, and thermal conductivity \cite{Yamashita09a}) indicate a
gapped state, the specific heat is linear in T at low temperature
consistent with a gapless energy spectrum \cite{Yamashita08a}.

\subsection{EtMe$_3$Sb[Pd(dmit)$_2$]$_2$}

dmit-Sb is similar in many respects to $\kappa$-CN: in terms of dimer
units, the frustration $t^\prime/t$ and effective $J$ for the two
materials are similar.  No apparent magnetic ordering occurs down to
mK temperature \cite{Itou10a}.  Frequency-dependent dielectric
response similar to that of $\kappa$-CN is found at low temperatures
\cite{AbdelJawad11a}. In the related material
EtMe$_3$P[Pd(dmit)$_2$]$_2$ described as having a VBS ground state
\cite{Tamura09a}, peak splitting below 25 K in the infrared optical
response indicates that the VBS state is charge ordered
\cite{Yamamoto11a}. The charge order patterns in EtMe$_3$P are in
complete agreement with those in the PEC \cite{Li10a,Dayal11a}.  In
dmit-Sb, {\it the same IR peaks are not split but broadened},
prompting the investigators to suggest fluctuating charge and bond
order \cite{Yamamoto11a}.  Similar to the 6 K transition in
$\kappa$-CN, a transition to a broken-symmetry state appears to occur
in dmit-Sb at $T\sim$1 K.  The $^{13}$C NMR relaxation rate $1/T_1$
decreases below 1 K, but with a different temperature dependence than
in $\kappa$-CN \cite{Shimizu06a,Itou10a,Itou11a}.  Thermal
conductivity measurements in weak magnetic fields suggest gapless
mobile excitations, but the steep increase of the thermal conductivity
when $H>$2 T shows the presence of gapped spin excitations
\cite{Yamashita10a}.

\subsection{Conclusion}

Within the effective $\frac{1}{2}$-filled model little difference
would be expected between $\kappa$-CN and dmit-Sb as both have similar
values of inter-dimer $t^\prime/t$.  While both materials display a
mix of gapped and gapless behavior, there are significant differences
between the two such as the temperature dependence of the $^{13}$C
$1/T_1$, which is $\propto T^{\frac{3}{2}}$ in $\kappa$-CN
\cite{Shimizu06a} but $\propto T^2$ in dmit-Sb \cite{Itou10a,Itou11a}.
These and other differences suggest that a $\frac{1}{4}$-filled
description is essential at low temperature.  We have suggested that
the ground states of $\kappa$-CN and dmit-Sb are PECs
\cite{Li10a,Dayal11a}. In such a picture the state below 6 K in
$\kappa$-CN and 1 K in dmit-Sb would be a charge disproportionated
spin gap state with fluctuating short-range order.  $\mu$SR gives
clear evidence for a small but finite spin gap from the
ground state in $\kappa$-CN\cite{Pratt11a}, in agreement with our
prediction.  A second transition at higher field\cite{Pratt11a} is,
however, currently not understood.  An alternate plausible model 
 is the non-magnetic insulator phase found in the
$\frac{1}{2}$-filled model, which however has no spin gap
\cite{Mizusaki06a}.

 A paired state
like the PEC can also explain the observed mix of gapped and gapless
behavior.  A near degeneracy of spin-singlet configurations can give a
linear specific heat. This degeneracy is particularly expected in
$\kappa$-CN, where the lattice structure suggests many nearly
equivalent ways to construct the spin-singlet bonds of the PEC
\cite{Li10a}. On the other hand, magnetic excitations require breaking
singlet bonds and will be gapped. A similar combination of gapless
specific heat but gapped magnetic susceptibility and conductivity is
found in vanadium bronzes \cite{Chakraverty78a}. It was suggested
singlet inter-site bipolarons could explain these features---
tunneling of the carriers composing the bipolarons between
nearly-equivalent lattice sites gives a finite degeneracy and linear
specific heat, while magnetic susceptibility is gapped because of the
pairing gap of the bipolarons \cite{Chakraverty78a}.

The suggestion that a paired state can describe the low temperature
properties of $\kappa$-CN and dmit-Sb also gives a possible mechanism
of superconducting pairing
\cite{Mazumdar08a,Li10a,Dayal11a,iscom2011-sm}.  In the picture,
spin-singlet pairs formed in real space would form a superconducting
paired-electron liquid \cite{Schafroth55a}. Within a simplified model
where pairs of sites in the $\frac{1}{4}$-filled model are replaced by
single sites in an effective $\frac{1}{2}$-filled model with
attractive ($-U$) interactions, a charge-order-to-SC transition occurs
in the presence of lattice frustration \cite{Mazumdar08a}.  We
emphasize that our proposed mechanism is different from the mean-field
theory of charge fluctuation driven SC \cite{Merino01a}.  Strong
correlation, $\frac{1}{4}$-filling, and lattice frustration are also
characteristic of unconventional SC in other families of materials,
including the layered cobaltates, spinels, and fullerides
\cite{iscom2011-sm}.

\begin{acknowledgement}
This work was supported by the US Department of Energy grant
DE-FG02-06ER46315. 
\end{acknowledgement}

\providecommand{\WileyBibTextsc}{}
\let\textsc\WileyBibTextsc
\providecommand{\othercit}{}
\providecommand{\jr}[1]{#1}
\providecommand{\etal}{~et~al.}


\begin{thebibliography}{[10]}

\bibitem{Balents10a}
 \textsc{L.~Balents},
 \jr{Nature} \textbf{464}, 199 (2010).


\bibitem{Kanoda11a}
 \textsc{K.~Kanoda} and  \textsc{R.~Kato},
 \jr{Annu. Rev. Condens. Matter Phys.} \textbf{2011}, 18.1 (2011).


\bibitem{Dressel11a}
 \textsc{M.~Dressel},
 \jr{J. Phys.: Condens. Matter} \textbf{23}, 293201 (2011).


\bibitem{Kato04a}
 \textsc{R.~Kato},
 \jr{Chem. Rev.} \textbf{104}, 5319 (2004).


\bibitem{Powell11a}
 \textsc{B.\,J. Powell} and  \textsc{R.\,H. McKenzie},
 \jr{Rep. Progr. Phys.} \textbf{74}, 056501 (2011).


\bibitem{Clay08a}
 \textsc{R.\,T. Clay},  \textsc{H.~Li},  and  \textsc{S.~Mazumdar},
 \jr{Phys.\ Rev.\ Lett.} \textbf{101}, 166403 (2008).


\bibitem{Tocchio09a}
 \textsc{L.\,F. Tocchio},  \textsc{A.~Parola},  \textsc{C.~Gros},  and
  \textsc{F.~Becca},
 \jr{Phys.\ Rev.\ B} \textbf{80}, 064419 (2009).


\bibitem{Kashima01b}
 \textsc{T.~Kashima} and  \textsc{M.~Imada},
 \jr{J.\ Phys.\ Soc.\ Jpn.} \textbf{70}, 2287 (2001).


\othercit
\bibitem{Dayal11b}
 \textsc{S.~Dayal},  \textsc{R.\,T. Clay},  and  \textsc{S.~Mazumdar},
in preparation.


\bibitem{Li10a}
 \textsc{H.~Li},  \textsc{R.\,T. Clay},  and  \textsc{S.~Mazumdar},
 \jr{J. Phys.: Condens. Matter} \textbf{22}, 272201 (2010).


\bibitem{Dayal11a}
 \textsc{S.~Dayal},  \textsc{R.\,T. Clay},  \textsc{H.~Li},  and
  \textsc{S.~Mazumdar},
 \jr{Phys.\ Rev.\ B} \textbf{83}, 245106 (2011).


\bibitem{Komatsu96a}
 \textsc{T.~Komatsu},  \textsc{N.~Matsukawa},  \textsc{T.~Inoue},  and
  \textsc{G.~Saito},
 \jr{J.\ Phys.\ Soc.\ Jpn.} \textbf{65}, 1340 (1996).


\bibitem{Shimizu07a}
 \textsc{Y.~Shimizu} \etal{},
 \jr{Phys.\ Rev.\ Lett.} \textbf{99}, 256403 (2007).


\bibitem{Shimizu03a}
 \textsc{Y.~Shimizu} \etal{},
 \jr{Phys.\ Rev.\ Lett.} \textbf{91}, 107001 (2003).


\bibitem{Shimizu06a}
 \textsc{Y.~Shimizu} \etal{},
 \jr{Phys.\ Rev.\ B} \textbf{73}, 140407 (2006).


\bibitem{Yamashita08a}
 \textsc{S.~Yamashita} \etal{},
 \jr{Nature Phys.} \textbf{4}, 459 (2008).


\bibitem{Yamashita09a}
 \textsc{M.~Yamashita} \etal{},
 \jr{Nature Phys.} \textbf{5}, 44 (2009).


\bibitem{AbdelJawad10a}
 \textsc{M.~Abdel-Jawad} \etal{},
 \jr{Phys.\ Rev.\ B} \textbf{82}, 125119 (2010).


\bibitem{Manna10a}
 \textsc{R.~Manna} \etal{},
 \jr{Phys.\ Rev.\ Lett.} \textbf{104}, 016403 (2010).


\bibitem{Pratt11a}
 \textsc{F.\,L. Pratt} \etal{},
 \jr{Nature} \textbf{471}, 612 (2011).


\bibitem{Itou10a}
 \textsc{T.~Itou},  \textsc{A.~Oyamada},  \textsc{S.~Maegawa},  and
  \textsc{R.~Kato},
 \jr{Nature Phys.} \textbf{6}, 673 (2010).


\othercit
\bibitem{AbdelJawad11a}
M. Abdel-Jawad et al., present conference.


\bibitem{Tamura09a}
 \textsc{M.~Tamura} and  \textsc{R.~Kato},
 \jr{Sci. Technol. Adv. Mater.} \textbf{10}, 024304 (2009).


\othercit
\bibitem{Yamamoto11a}
T. Yamamoto et al., present conference.


\bibitem{Itou11a}
 \textsc{T.~Itou} \etal{},
 \jr{Phys. Rev. B} \textbf{84}, 094405 (2011).


\bibitem{Yamashita10a}
 \textsc{M.~Yamashita} \etal{},
 \jr{Science} \textbf{328}, 1246 (2010).


\bibitem{Mizusaki06a}
 \textsc{T.~Mizusaki} and  \textsc{M.~Imada},
 \jr{Phys.\ Rev.\ B} \textbf{74}, 014421 (2006).


\bibitem{Chakraverty78a}
 \textsc{B.\,K. Chakraverty},  \textsc{M.\,J. Sienko},  and
  \textsc{J.~Bonnerot},
 \jr{Phys.\ Rev.\ B} \textbf{17}, 3781 (1978).


\bibitem{Mazumdar08a}
 \textsc{S.~Mazumdar} and  \textsc{R.\,T. Clay},
 \jr{Phys.\ Rev.\ B} \textbf{77}, 180515(R) (2008).


\othercit
\bibitem{iscom2011-sm}
S. Mazumdar and R.~T. Clay, see present proceedings.


\bibitem{Schafroth55a}
 \textsc{M.\,R. Schafroth},
 \jr{Phys. Rev.} \textbf{100}, 463 (1955).


\bibitem{Merino01a}
 \textsc{J.~Merino} and  \textsc{R.\,H. McKenzie},
 \jr{Phys.\ Rev.\ Lett.} \textbf{87}, 237002 (2001).


\end{thebibliography}
\end{document}